\pdfoutput=1
\documentclass[12pt,a4paper]{article}

\usepackage{ifthen} 
\usepackage{verbatim}
\newboolean{pdflatex}
\setboolean{pdflatex}{true} 

\newboolean{articletitles}
\setboolean{articletitles}{true} 

\newboolean{uprightparticles}
\setboolean{uprightparticles}{false} 

\usepackage{booktabs}

\newcommand{\Tccbar}{\ensuremath{T_{\ccbar1}}\xspace}
\newcommand{\Zcm}{\ensuremath{T_{\ccbar1}(4430)^-}\xspace}

\def\paperauthors{LHCb collaboration} 
\def\paperasciititle{Observation of the Tcc(4430) state in B0->D*DK decays} 
\def\papertitle{Observation of the \Zcm state in \decay{\Bz}{\Dstarm \Dz \Kp} decays} 
\def\paperkeywords{{High Energy Physics}, {LHCb}} 
\def\papercopyright{\the\year\ CERN for the benefit of the LHCb collaboration} 
\def\paperlicence{CC BY 4.0 licence}
\def\paperlicenceurl{https://creativecommons.org/licenses/by/4.0/}

\newif\ifEnableSectionTOCLinks
\EnableSectionTOCLinksfalse 

\usepackage[top=1in, bottom=1.25in, left=1in, right=1in]{geometry}

\usepackage{microtype}
\usepackage{lineno}  
\usepackage{xspace} 
\usepackage{caption} 

\usepackage{graphicx}  
\usepackage{color}
\usepackage{colortbl}
\graphicspath{{./figs/}} 

\usepackage{amsmath} 
\usepackage{amssymb}
\usepackage{amsfonts}
\usepackage{upgreek} 

\usepackage[normalem]{ulem} 

\newcommand*\patchAmsMathEnvironmentForLineno[1]{%
\expandafter\let\csname old#1\expandafter\endcsname\csname #1\endcsname
\expandafter\let\csname oldend#1\expandafter\endcsname\csname
end#1\endcsname
 \renewenvironment{#1}%
   {\linenomath\csname old#1\endcsname}%
   {\csname oldend#1\endcsname\endlinenomath}%
}
\newcommand*\patchBothAmsMathEnvironmentsForLineno[1]{%
  \patchAmsMathEnvironmentForLineno{#1}%
  \patchAmsMathEnvironmentForLineno{#1*}%
}
\AtBeginDocument{%
\patchBothAmsMathEnvironmentsForLineno{equation}%
\patchBothAmsMathEnvironmentsForLineno{align}%
\patchBothAmsMathEnvironmentsForLineno{flalign}%
\patchBothAmsMathEnvironmentsForLineno{alignat}%
\patchBothAmsMathEnvironmentsForLineno{gather}%
\patchBothAmsMathEnvironmentsForLineno{multline}%
\patchBothAmsMathEnvironmentsForLineno{eqnarray}%
}

\usepackage[pdftex,
            pdfauthor={\paperauthors},
            pdftitle={\paperasciititle},
            pdfkeywords={\paperkeywords}]{hyperref}
\usepackage{hyperxmp}
\hypersetup{
    pdfcopyright={Copyright (C) \papercopyright},
    pdflicenseurl={\paperlicenceurl}
}

\usepackage[colorinlistoftodos,textsize=scriptsize]{todonotes}

\usepackage[bottom,flushmargin,hang,multiple]{footmisc}

\usepackage[all]{hypcap} 

\usepackage{xspace} 
\usepackage{upgreek}

\def\lhcb   {\mbox{LHCb}\xspace}

\def\babar  {\mbox{BaBar}\xspace}
\def\belle  {\mbox{Belle}\xspace}

\def\MagUp {\mbox{\em Mag\kern -0.05em Up}\xspace}

\ifdefined\ifuprightparticles
\else
\newboolean{uprightparticles}
\setboolean{uprightparticles}{false} 
\fi

\ifthenelse{\boolean{uprightparticles}}%
{

 \def\Ppi         {\ensuremath{\uppi}\xspace}

 \def\Ppsi        {\ensuremath{\uppsi}\xspace}

 \def\PDelta      {\ensuremath{\Delta}\xspace}                 
 \def\PXi         {\ensuremath{\Xi}\xspace}                 
 \def\PLambda     {\ensuremath{\Lambda}\xspace}                 
 \def\PSigma      {\ensuremath{\Sigma}\xspace}                 
 \def\POmega      {\ensuremath{\Omega}\xspace}                 
 \def\PUpsilon    {\ensuremath{\Upsilon}\xspace}
 \let\oldPi\Pi
 \def\PPi         {\ensuremath{\oldPi}\xspace}

 \def\PB      {\ensuremath{\mathrm{B}}\xspace}                 
 \def\PD      {\ensuremath{\mathrm{D}}\xspace}                 
 \def\PJ      {\ensuremath{\mathrm{J}}\xspace}                 
 \def\PK      {\ensuremath{\mathrm{K}}\xspace}                 

 \def\Pc      {\ensuremath{\mathrm{c}}\xspace}

 \def\Pq      {\ensuremath{\mathrm{q}}\xspace}                 
 \def\Ps      {\ensuremath{\mathrm{s}}\xspace}

 \def\thebaroffset{0.0em}
}
{

 \def\Ppi         {\ensuremath{\pi}\xspace}

 \def\Ppsi        {\ensuremath{\psi}\xspace}                 
                  
 \mathchardef\PDelta="7101
 \mathchardef\PXi="7104
 \mathchardef\PLambda="7103
 \mathchardef\PSigma="7106
 \mathchardef\POmega="710A
 \mathchardef\PUpsilon="7107
 \mathchardef\PPi="7105
 \def\PB      {\ensuremath{B}\xspace}                 
 \def\PD      {\ensuremath{D}\xspace}                 
 \def\PJ      {\ensuremath{J}\xspace}                 
 \def\PK      {\ensuremath{K}\xspace}                 

 \def\Pc      {\ensuremath{c}\xspace}

 \def\Pq      {\ensuremath{q}\xspace}                 
 \def\Ps      {\ensuremath{s}\xspace}

 \def\thebaroffset{0.18em}
}
\newcommand{\offsetoverline}[2][\thebaroffset]{\kern #1\overline{\kern -#1 #2}}%

\makeatletter
\ifcase \@ptsize \relax
  \newcommand{\miniscule}{\@setfontsize\miniscule{4}{5}}
\or
  \newcommand{\miniscule}{\@setfontsize\miniscule{5}{6}}
\or
  \newcommand{\miniscule}{\@setfontsize\miniscule{5}{6}}
\fi
\makeatother

\DeclareRobustCommand{\optbar}[1]{\shortstack{{\miniscule (\rule[.5ex]{1.25em}{.18mm})}
  \\ [-.7ex] $#1$}}

\def\quark     {{\ensuremath{\Pq}}\xspace}
\def\quarkbar  {{\ensuremath{\overline \quark}}\xspace}
\def\qqbar     {{\ensuremath{\quark\quarkbar}}\xspace}

\def\squark    {{\ensuremath{\Ps}}\xspace}
\def\squarkbar {{\ensuremath{\overline \squark}}\xspace}

\def\cquark    {{\ensuremath{\Pc}}\xspace}
\def\cquarkbar {{\ensuremath{\overline \cquark}}\xspace}
\def\ccbar     {{\ensuremath{\cquark\cquarkbar}}\xspace}

\def\pion   {{\ensuremath{\Ppi}}\xspace}

\def\pip    {{\ensuremath{\pion^+}}\xspace}
\def\pim    {{\ensuremath{\pion^-}}\xspace}
\def\pipm   {{\ensuremath{\pion^\pm}}\xspace}
\def\pimp   {{\ensuremath{\pion^\mp}}\xspace}

\def\kaon    {{\ensuremath{\PK}}\xspace}

\def\KorKbar {\kern \thebaroffset\optbar{\kern -\thebaroffset \PK}{}\xspace}

\def\Kp      {{\ensuremath{\kaon^+}}\xspace}

\def\Kpm     {{\ensuremath{\kaon^\pm}}\xspace}

\def\KS      {{\ensuremath{\kaon^0_{\mathrm{S}}}}\xspace}

\def\Dbar    {{\ensuremath{\offsetoverline{\PD}}}\xspace}
\def\D       {{\ensuremath{\PD}}\xspace}

\def\DorDbar {\kern \thebaroffset\optbar{\kern -\thebaroffset \PD}\xspace}
\def\Dz      {{\ensuremath{\D^0}}\xspace}
\def\Dzb     {{\ensuremath{\Dbar{}^0}}\xspace}
\def\Dp      {{\ensuremath{\D^+}}\xspace}
\def\Dm      {{\ensuremath{\D^-}}\xspace}

\def\DpDm    {\ensuremath{\Dp {\kern -0.16em \Dm}}\xspace}
\def\Dstar   {{\ensuremath{\D^*}}\xspace}
\def\Dstarb  {{\ensuremath{\Dbar{}^*}}\xspace}

\def\Dstarm  {{\ensuremath{\D^{*-}}}\xspace}

\def\Dsm     {{\ensuremath{\D^-_\squark}}\xspace}

\def\Dssm    {{\ensuremath{\D^{*-}_\squark}}\xspace}

\def\B       {{\ensuremath{\PB}}\xspace}

\def\BorBbar {\kern \thebaroffset\optbar{\kern -\thebaroffset \PB}\xspace}
\def\Bz      {{\ensuremath{\B^0}}\xspace}

\def\Bd      {{\ensuremath{\B^0}}\xspace}

\def\BdorBdbar {\kern \thebaroffset\optbar{\kern -\thebaroffset \Bd}\xspace}
\def\Bu      {{\ensuremath{\B^+}}\xspace}

\def\Bp      {{\ensuremath{\Bu}}\xspace}

\def\Bs      {{\ensuremath{\B^0_\squark}}\xspace}

\def\BsorBsbar {\kern \thebaroffset\optbar{\kern -\thebaroffset \Bs}\xspace}

\def\jpsi     {{\ensuremath{{\PJ\mskip -3mu/\mskip -2mu\Ppsi}}}\xspace}
\def\psitwos  {{\ensuremath{\Ppsi{(2S)}}}\xspace}

\def\Y#1S{\ensuremath{\PUpsilon{(#1S)}}\xspace}

\def\LorLbar     {\kern \thebaroffset\optbar{\kern -\thebaroffset \PLambda}\xspace}

\newcommand{\decay}[2]{\ensuremath{\mathinner{#1\!\to #2}}\xspace}

\def\to                 {\ensuremath{\rightarrow}\xspace}

\def\AT#1     {\ensuremath{A_{\mathrm{T}}^{#1}}\xspace}           

\def\C#1      {\ensuremath{\mathcal{C}_{#1}}\xspace}                       
\def\Cp#1     {\ensuremath{\mathcal{C}_{#1}^{'}}\xspace}                    
\def\Ceff#1   {\ensuremath{\mathcal{C}_{#1}^{\mathrm{(eff)}}}\xspace}        
\def\Cpeff#1  {\ensuremath{\mathcal{C}_{#1}^{'\mathrm{(eff)}}}\xspace}       
\def\Ope#1    {\ensuremath{\mathcal{O}_{#1}}\xspace}                       
\def\Opep#1   {\ensuremath{\mathcal{O}_{#1}^{'}}\xspace}                    

\newcommand{\aunit}[1]{\ensuremath{\text{\,#1}}}       

\newcommand{\tev}{\aunit{Te\kern -0.1em V}\xspace}
\newcommand{\gev}{\aunit{Ge\kern -0.1em V}\xspace}
\newcommand{\mev}{\aunit{Me\kern -0.1em V}\xspace}
\newcommand{\kev}{\aunit{ke\kern -0.1em V}\xspace}
\newcommand{\ev}{\aunit{e\kern -0.1em V}\xspace}
 
\newcommand{\mevc}{\ensuremath{\aunit{Me\kern -0.1em V\!/}c}\xspace}
\newcommand{\gevc}{\ensuremath{\aunit{Ge\kern -0.1em V\!/}c}\xspace}
\newcommand{\mevcc}{\ensuremath{\aunit{Me\kern -0.1em V\!/}c^2}\xspace}
\newcommand{\gevcc}{\ensuremath{\aunit{Ge\kern -0.1em V\!/}c^2}\xspace}

\def\fb   {\ensuremath{\aunit{fb}}\xspace}
\def\invfb   {\ensuremath{\fb^{-1}}\xspace}

\newcommand{\chisq}{\ensuremath{\chi^2}\xspace}

\def\gsim{{~\raise.15em\hbox{$>$}\kern-.85em
          \lower.35em\hbox{$\sim$}~}\xspace}
\def\lsim{{~\raise.15em\hbox{$<$}\kern-.85em
          \lower.35em\hbox{$\sim$}~}\xspace}

\def\tell1  {TELL1\xspace}
\def\ukl1   {UKL1\xspace}

\newcommand{\lhcborcid}[1]{\href{https://orcid.org/#1}{\hspace*{0.1em}\raisebox{-0.45ex}{\includegraphics[width=1em]{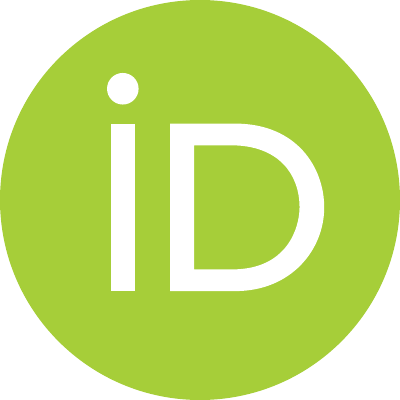}}}}

\hypersetup{
  colorlinks   = true, 
  urlcolor     = blue, 
  linkcolor    = blue, 
  citecolor    = red   
}

\ifEnableSectionTOCLinks
    \usepackage[explicit]{titlesec} 
    
    \let\oldcontentsline\contentsline
    \renewcommand

    \titleformat{\section}{\normalfont\Large\bf}{\hyperlink{tocsection.\thesection}{{\thesection} \parbox[t]{\dimexpr\textwidth-1pc}{#1}}}{1pc}{}

    \titleformat{\subsection}{\normalfont\bf}{\hyperlink{tocsubsection.\thesubsection}{{\thesubsection} \parbox[t]{\dimexpr\textwidth-1pc}{#1}}}{1pc}{}

    \titleformat{name=\section,numberless}[display]{}{}{0pt}{\normalfont\Huge\bfseries #1}
\fi

\usepackage{cite} 
\usepackage{LHCb/mciteplus}
\makeatletter
\g@addto@macro\bfseries{\boldmath}
\makeatother

\usepackage{longtable} 

\begin{document}

\renewcommand{\thefootnote}{\fnsymbol{footnote}}
\setcounter{footnote}{1}


\begin{titlepage}
\pagenumbering{roman}

\vspace*{-1.5cm}
\centerline{\large EUROPEAN ORGANIZATION FOR NUCLEAR RESEARCH (CERN)}
\vspace*{1.5cm}
\noindent
\begin{tabular*}{\linewidth}{lc@{\extracolsep{\fill}}r@{\extracolsep{0pt}}}
\ifthenelse{\boolean{pdflatex}}
{\vspace*{-1.5cm}\mbox{\!\!\!\includegraphics[width=.14\textwidth]{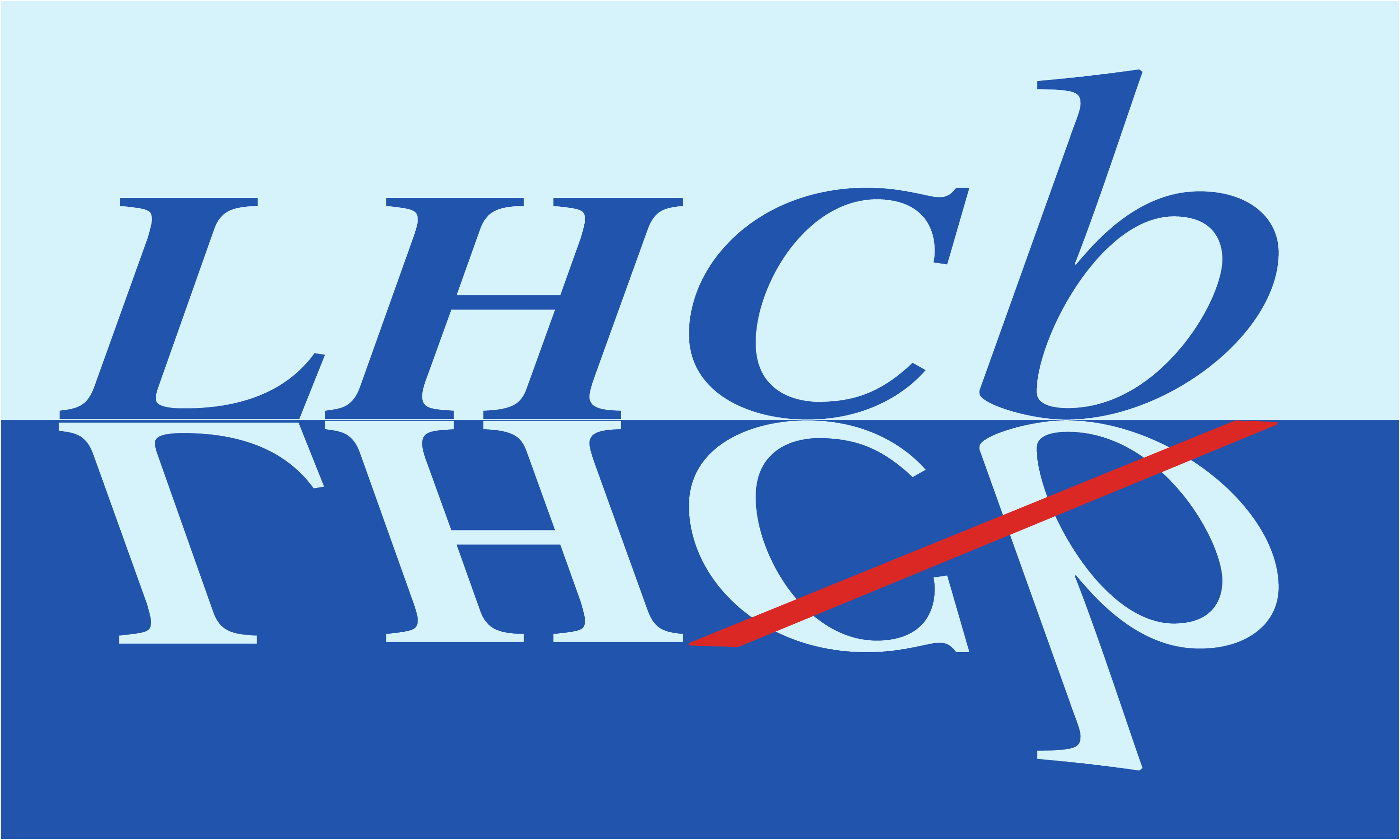}} & &}%
{\vspace*{-1.2cm}\mbox{\!\!\!\includegraphics[width=.12\textwidth]{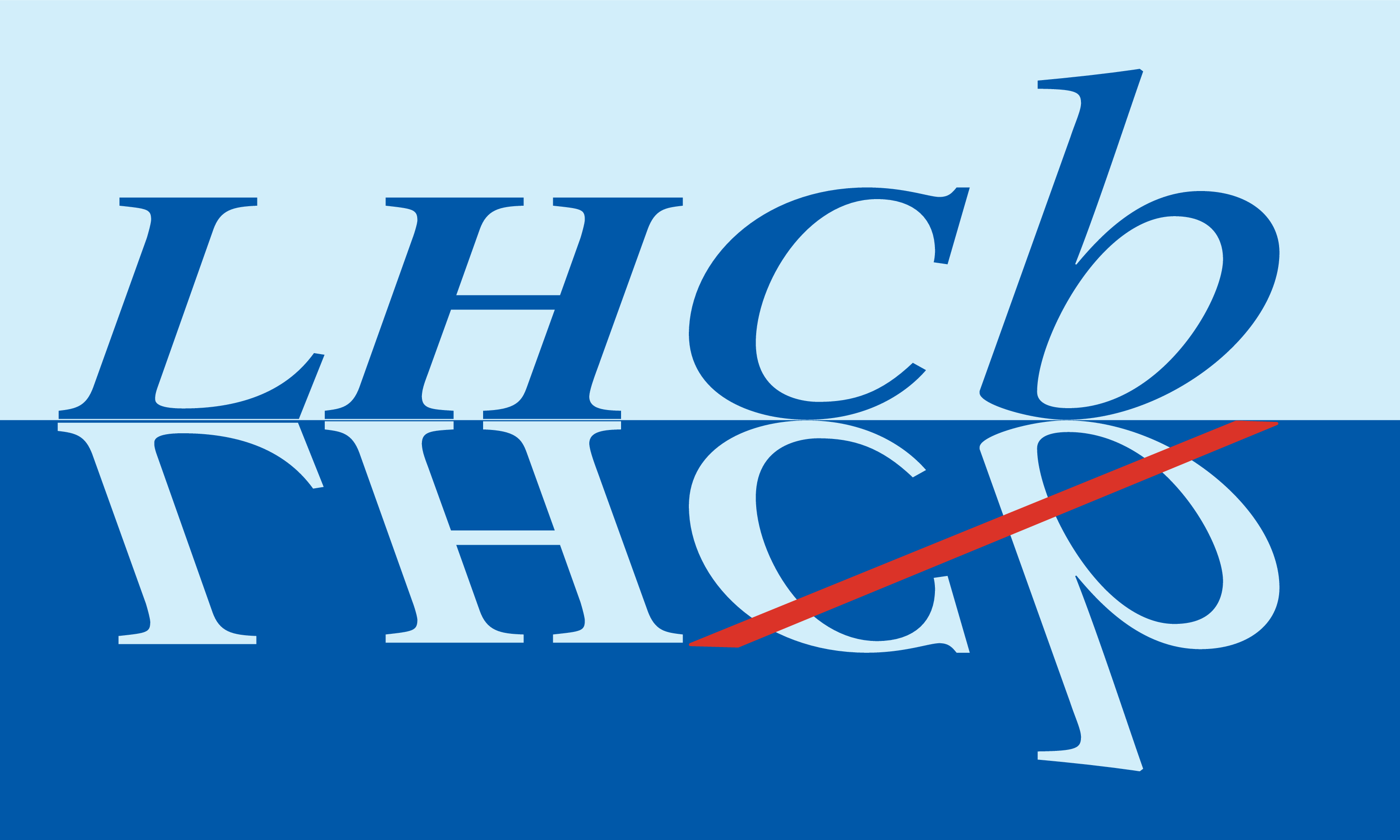}} & &}%
\\
 & & CERN-EP-2026-266 \\  
 & & LHCb-PAPER-2026-033 \\  
 & & October 25, 2026 \\ 
 & & \\
\end{tabular*}

\vspace*{4.0cm}

{\normalfont\bfseries\boldmath\huge
\begin{center}
  \papertitle 
\end{center}
}

\vspace*{2.0cm}

\begin{center}
\paperauthors\footnote{Authors are listed at the end of this Letter.}
\end{center}

\vspace{\fill}

\begin{abstract}
    \noindent
    A study of resonant structures in $B^0 \to \Dstarm \Dz \Kp$ decays is presented.
    The analysis uses proton-proton collision data collected by the LHCb experiment at center-of-mass energies of $\sqrt{s} = 7$, 8, and 13\tev, corresponding to an integrated luminosity of 9\invfb.
    The analysis demonstrates that the existing known excited $D_s^+$ resonances in the $\Dz\Kp$ spectrum are insufficient to  describe the data.
    The inclusion of the tetraquark state $T_{c\bar{c}1}(4430)^-$ improves the description of the data with a statistical significance exceeding $5\sigma$. This corresponds to the first observation of the $T_{c\bar{c}1}(4430)^- \to D^{*-} D^0$ decay.
    The measured properties of the $T_{c\bar{c}1}(4430)^-$ state are consistent with those previously measured in its decay to the $\psi(2S)\pim$ final state.

\end{abstract}

\vspace*{2.0cm}

\begin{center}
  Submitted to
  Phys.~Rev.~Lett.
\end{center}

\vspace{\fill}

{\footnotesize 
\centerline{\copyright~\papercopyright. \href{\paperlicenceurl}{\paperlicence}.}}
\vspace*{2mm}

\end{titlepage}


\newpage
\setcounter{page}{2}
\mbox{~}
%
%
%
%


\renewcommand{\thefootnote}{\arabic{footnote}}
\setcounter{footnote}{0}


\cleardoublepage


\pagestyle{plain} 
\setcounter{page}{1}
\pagenumbering{arabic}



The study of exotic hadrons provides a crucial testing ground for Quantum Chromodynamics (QCD) in its nonperturbative regime.
Charged charmonium-like states are particularly interesting because their minimal quark content of $\ccbar\qqbar^\prime$ makes them unambiguously exotic hadrons. Proposed interpretations of their internal structure include molecular states~\cite{Guo_2018} and compact tetraquarks~\cite{Esposito_2017}. Detailed studies are needed to distinguish between these scenarios.

The \Zcm state was first reported by the \belle collaboration in \decay{\Bz}{\psitwos \Kp \pim} decays~\cite{PhysRevLett.100.142001}, though not confirmed in a subsequent search by the \babar collaboration~\cite{Babar:2008bxw}. Later analyses by the \belle collaboration confirmed the initial claim with high significance and performed the first measurement of the \Zcm spin and parity~\cite{Belle:2009lvn,Belle:2013shl}.
Subsequently, the \lhcb collaboration has firmly established its resonant nature and its spin-parity to be ${J^P = 1^+}$ through a comprehensive amplitude analysis of \decay{\Bz}{\psitwos \Kp \pim} decays~\cite{LHCb-PAPER-2014-014}.
An \lhcb model-independent analysis of the same process~\cite{LHCb-PAPER-2015-038} and the subsequent studies of new $B$-decay channels such as the \decay{\Bp}{\psitwos \Kp \pip \pim}~\cite{LHCb-PAPER-2024-014} and the isospin-related \decay{\Bp}{\psitwos \KS \pip} decays~\cite{LHCb-PAPER-2025-039} confirmed the observation of the state. 
While the Belle collaboration found evidence for the \decay{\Zcm}{\jpsi\pim}~\cite{Belle:2014nuw} decay, a subsequent \lhcb analysis using a model-independent approach~\cite{LHCb-PAPER-2018-043} found clear exotic contributions in the mass region around~4.2\gev  and~4.6\gev, but reached no conclusion on \Zcm.
Natural units with $\hbar=c=1$ are used throughout.
The spectrum of charged exotic hadrons in this region is further complicated by the $\Tccbar(4200)^-$ state, which has also been observed in both \decay{\Bz}{\psitwos \Kp \pim} and \decay{\Bz}{\jpsi \Kp \pim} decays~\cite{LHCb-PAPER-2014-014, Belle:2014nuw, LHCb-PAPER-2018-043, LHCb-PAPER-2024-014}.

The structure of the \Zcm state has been interpreted through various models, most notably as a $\Dstar\Dbar_1^{(\prime)}$ molecular state, given its proximity to this threshold, or as a radial excitation of the $\Tccbar(3900)^-$ state in a compact diquark-antidiquark picture, as suggested by the $\psitwos$--$\jpsi$ mass-splitting analogy~\cite{4430_2, 4430_5, 4430_6, PhysRevD.77.034003, PhysRevD.79.034026, PhysRevD.89.114010, branch1, branch2}.
Alternative explanations include cross-channel effects enhanced by a triangle singularity from open-charm states~\cite{Pakhlov:2011xj, Pakhlov:2014qva, Nakamura:2019btl, Cheng:2026cgo}, which are also used to interpret the $\Tccbar(4200)^-$ state.
The relative strength of its open-charm and hidden-charm decays serves as a critical discriminator between these models\cite{branch2}. The absence to date of any observed open-charm decay mode makes its experimental measurement a paramount goal for understanding the nature of the \Zcm state.
The \decay{\Bz}{\Dstarm \Dz \Kp} decay is an ideal platform for performing this study. 
It also offers an opportunity to search for the $\Tccbar(3900)^-$ state, first observed in the $\D\Dstarb$
channel~\cite{Zc3900_BES14,Zc3900_BES15}, and the $T^*_{\cquarkbar\squarkbar1}(2900)^0$ state, first observed in the $\Dm\Kp$ channel~\cite{LHCb-PAPER-2020-024,LHCb-PAPER-2020-025}, through the \Dstarm\Dz and \Dstarm\Kp subsystems, respectively.
Observing these exotic states in other decay modes would enable critical tests of tetraquark dynamics.
This would also provide valuable input for understanding the excited \Dsm spectrum. 


This Letter reports the results of an amplitude analysis of the \decay{\Bz}{\Dstarm \Dz \Kp} decay using proton-proton ($pp$) collision data collected by the \lhcb experiment, corresponding to an integrated luminosity of $3\invfb$ at center-of-mass energies $\sqrt{s}=7,~8\tev$, referred to as Run 1, and $6\invfb$ at $\sqrt{s}=13\tev$, referred to as Run 2. The inclusion of charge-conjugate particles is implied throughout this Letter.
The \lhcb detector is a single-arm forward spectrometer covering pseudorapidities  $2 < \eta < 5$ and is described in detail in Refs.~\cite{LHCb-DP-2008-001,LHCb-DP-2014-002}.
The detector acceptance and selection requirements are studied with simulated data produced using software packages described in Refs.~\cite{Sjostrand:2007gs, LHCb-PROC-2010-056, Lange:2001uf, davidson2015photos, Allison:2006ve, *Agostinelli:2002hh}.

The reconstruction and selection for \decay{\Bz}{\Dstarm \Dz \Kp} candidates follow the strategy established in Ref.~\cite{LHCb-PAPER-2020-006}, with $\decay{\Dstarm}{\Dzb}{\pim}$ and the \Dz and \Dzb mesons reconstructed through their decays to  the \Kpm\pimp or \Kpm\pimp\pimp\pipm final states. The case where both \Dz and \Dzb are reconstructed in the four-body mode is not considered due to its low yield.
The remaining three decay-mode combinations, in each of the two Run periods, form the six independent subsamples of the analysis.
To constitute a \Bz candidate, the reconstructed \Dstarm and \Dz candidates are combined with a track identified as a kaon.
A kinematic fit~\cite{Hulsbergen:2005pu} is applied, first constraining the \Dz, \Dzb and \Dstarm candidate masses to their known values~\cite{PDG2024} for the determination of the \Bz invariant mass, and subsequently adding a \Bz candidate mass constraint to obtain improved track four-momentum used in the amplitude fit.
A Boosted Decision Tree classifier~\cite{BBDT,TMVA4} that exploits topological, kinematic, and particle-identification variables is used to suppress combinatorial background, and the selection criteria are optimized to maximize the signal significance within the \Bz mass window $[5260, 5300]\mev$, taken as the signal region for the amplitude analysis.
Additional requirements on the invariant mass of the \Dz and \Dstarm candidates and their flight-distance significance from the \Bz decay vertex are applied to suppress backgrounds in which a \Bz decay to the same final-state particles is reconstructed with at most one intermediate charmed meson. After these selections, the non-double-charm background fraction averages approximately 1.5\% across all subsamples.
In events with multiple candidates, which occur in about 2\% of the sample, one candidate is chosen randomly.

\begin{figure}
    \centering
    \includegraphics[width=0.5\linewidth]{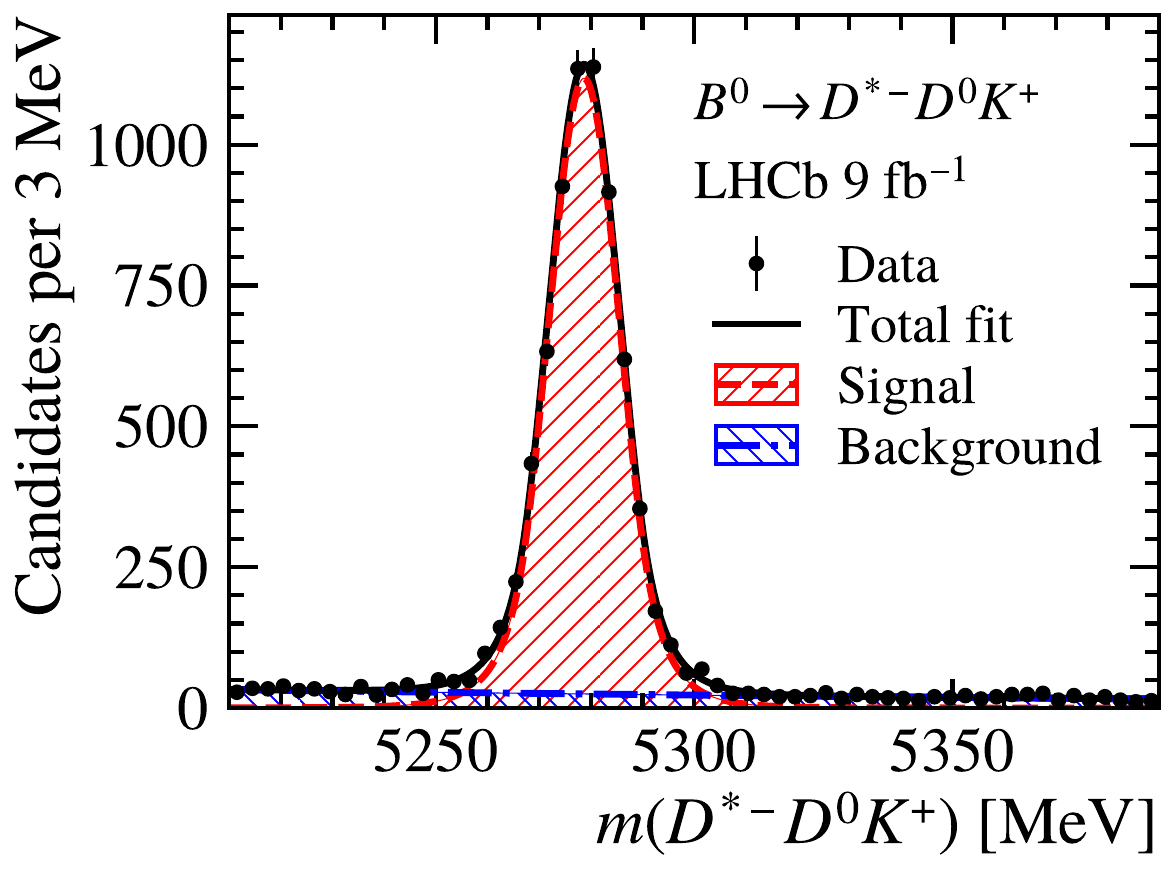}
    \caption{Invariant-mass distribution of \decay{\Bz}{\Dstarm \Dz \Kp} candidates, combining all \Dz final states and the full data sample. The black line shows the total fit, while the red and blue shaded areas represent the signal and background contributions, respectively.}
    \label{fig:Bmass}
\end{figure}

Signal and background yields are determined through a fit to the invariant-mass distributions of \Bz candidates in each subsample, where the signal is modeled with a double-sided Crystal Ball function~\cite{Skwarnicki:1986xj}, and the background with an exponential function. The fitted distributions for all data are shown in Fig.~\ref{fig:Bmass}. In total, 6912 candidates are used for the amplitude analysis, with a background fraction, $f_{\rm bkg}$, of approximately 5\%, which consists of both non-double-charm and combinatorial background.

\begin{figure}
    \centering
        \includegraphics[width=0.6\linewidth]{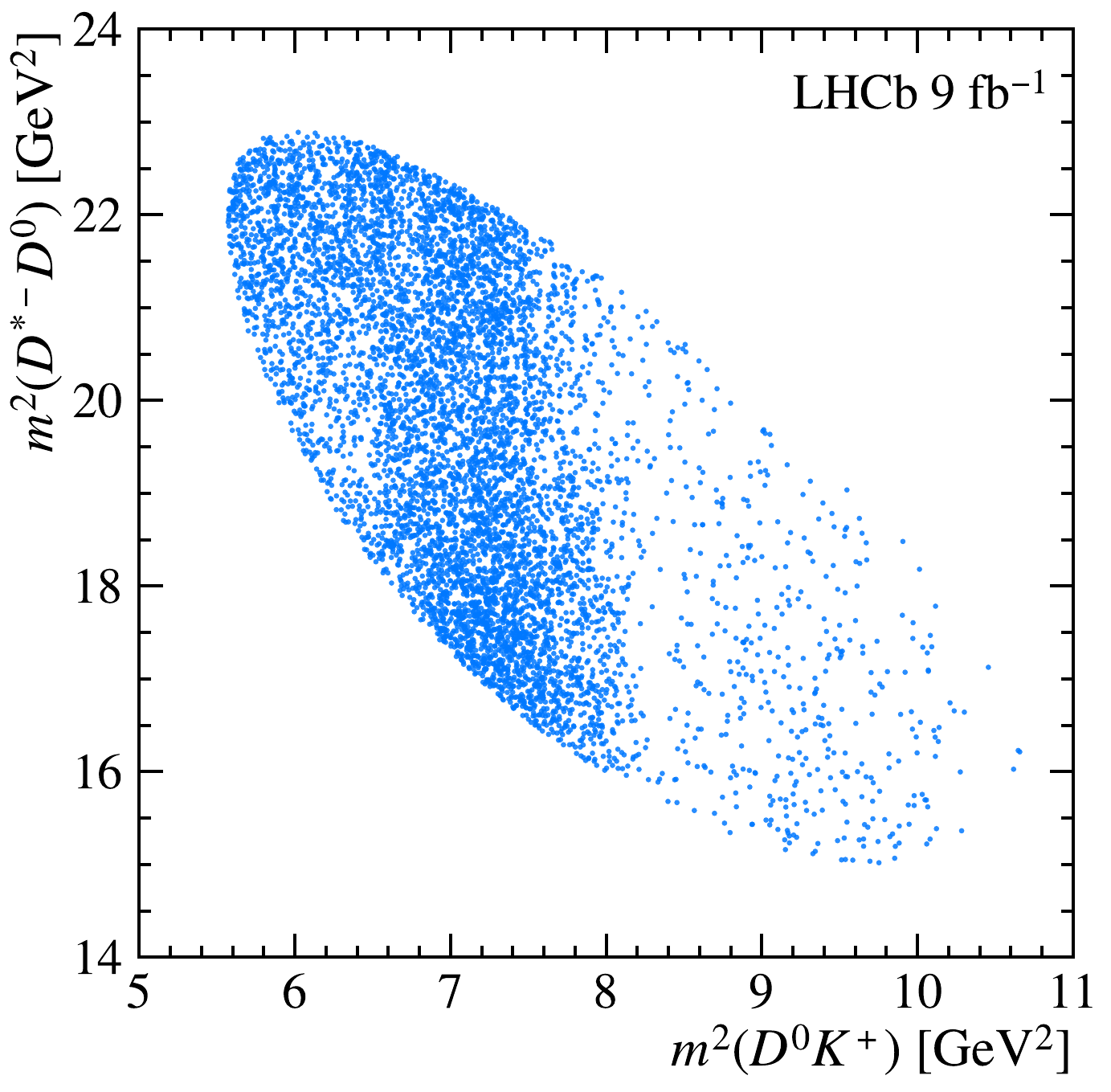}
    \caption{Dalitz plot for the selected \decay{\Bz}{\Dstarm \Dz \Kp} candidates.}
    \label{fig:dalitz}
\end{figure}


To determine the resonant contributions to the decays, an unbinned maximum-likelihood fit to the data is performed simultaneously across all six subsamples using the \texttt{TF-PWA} software~\cite{tf-pwa}. The negative log-likelihood (NLL) function for each subsample is defined as 
\begin{equation}
-\ln L = -\sum_{i} \ln \left[(1-f_{\rm bkg}) \frac{|\mathcal{A}(x_i)|^2}{\int |\mathcal{A}(x)|^2\epsilon(x) \mathrm{d} \Phi(x)} + f_{\rm bkg}\frac{B(x_i)}{\epsilon(x_i)} \right],
  \label{eqn:NLL}
\end{equation}
where $x_i$ represents the coordinates of the candidate $i$ in phase-space $\Phi$ and $\mathcal{A}(x)$ is the total signal decay amplitude, constructed as a coherent sum of the resonant and nonresonant contributions introduced below.
The quantities $\epsilon(x)$ and $B(x)$ denote the detection efficiency and the background probability density function.
The four-dimensional phase space is spanned by the two Dalitz-plot variables, $m^2(\Dstarm \Dz)$, $m^2(\Dz\Kp)$, and the two decay angles of the \decay{\Dstarm}{\Dzb}{\pim} decay, namely the cosine of the helicity angle for $\Dstarm$ and the azimuthal angle between the decay plane of \decay{\Dstarm}{\Dzb}{\pim} and that formed by $\Dz\Kp$ in the $\Bz$ rest frame.
Normalization factors are computed using simulated samples~\cite{Garmash_2005}, where  $\epsilon(x)$ is determined by applying identical selection criteria as for data and subsequently corrected for data-simulation discrepancies in particle identification~\cite{LHCb-DP-2018-001}, tracking~\cite{LHCb-DP-2013-002}, and trigger response~\cite{LHCb-DP-2012-004}. 
Candidates from the \Bz mass sideband region, defined as \mbox{$m(\Dstarm \Dz \Kp) \in[5320, 5390] \mev$}, are used to model $B(x)$, employing the kernel density estimation method~\cite{Poluektov_2015}, and $f_{\rm bkg}$ values are fixed to those obtained from the mass fits described above.
The terms $\epsilon(x)$, $B(x)$, and $f_{\rm bkg}$ are each determined independently for every subsample.
The total NLL for the simultaneous fit is the sum of the individual NLLs from all subsamples.

The Dalitz plot distribution for the selected \decay{\Bz}{\Dstarm \Dz \Kp} candidates is shown in Fig.~\ref{fig:dalitz}, where contributions from \Dssm resonances are visible as vertical bands.
To construct the signal amplitude $\mathcal{A}(x)$ entering
Eq.~\ref{eqn:NLL}, resonant and nonresonant contributions in all three
two-body subsystems are considered.
For conventional $\qqbar^\prime$ states, resonant structures are expected only in the \Dz\Kp invariant-mass spectrum. Based on current knowledge~\cite{PDG2024}, the \Dz\Kp spectrum includes four resonant contributions: $D_{\squark2}^*(2573)^+$, $D_{\squark1}^*(2700)^+$, $D_{\squark3}^*(2860)^+$, and $D_{\squark1}^*(2860)^+$.
Relativistic Breit--Wigner functions are used to describe the lineshapes of the $D_{\squark2}^*(2573)^+$ and $D_{\squark3}^*(2860)^+$ mesons, with their mass and width parameters fixed to the known values~\cite{PDG2024}.
The Blatt--Weisskopf barrier factor~\cite{VonHippel:1972fg} is included in all resonant amplitudes, with a radius parameter of $3.0\gev^{-1}$.
The decay \decay{\Bz}{\Dstarm D_{\squark3}^*(2860)^+} is included in the model, as the \decay{D_{\squark3}^*(2860)^+}{\Dz \Kp} decay has been observed in other analyses~\cite{LHCb-PAPER-2014-035,LHCb-PAPER-2014-036}. For this component, due to its minor role, only the lowest allowed orbital angular momentum ($D$-wave) is considered.
The two broad spin-1 states $D_{\squark1}^*(2700)^+$ and $D_{\squark1}^*(2860)^+$ largely overlap, prompting concerns about the validity of their description as a sum of relativistic Breit--Wigner functions.
Therefore, a quasi-model-independent (QMI) method~\cite{LHCB-PAPER-2016-026} and a \mbox{$K$-matrix} model~\cite{PDG2024_kmatrix} are adopted for the \Dz\Kp $J^P = 1^-$ components. The QMI description provides a better fit quality and is therefore chosen as the baseline model, while the \mbox{$K$-matrix} model is used to evaluate systematic uncertainties. In the QMI method, the contribution is divided into three separate partial waves corresponding to different orbital angular momenta between the \Dstarm and spin-1 components, each described by a cubic-spline interpolation.
The corresponding knot positions are listed in Table~\ref{tab:knots} in the End Matter.
The fit fractions for the $L = 0, 1, 2$ components are $(61.6 \pm 1.0)\%$, $(8.3 \pm 0.6)\%$ and $(1.7 \pm 0.2)\%$, respectively. 
The squared magnitude of the amplitude and phase of the $L=0$ contribution of the $J^P=1^-$ amplitude are shown as functions of $m(\Dz\Kp)$ in Fig.~\ref{fig:amp and phase} in the End Matter for three descriptions: the QMI spline fit; the $K$-matrix model; and a model comprising the sum of two relativistic Breit--Wigner amplitudes for the $D_{\squark1}^*(2700)^+$ and $D_{\squark1}^*(2860)^+$ states,  together with a uniform nonresonant $J^P=1^-$ contribution.
For the nonresonant component, only the $L=0$ partial wave is retained.
In the amplitude magnitude, the QMI and $K$-matrix results are broadly consistent, whereas the Breit--Wigner sum yields a narrower peak near 2.7\gev. In the phase, the $K$-matrix model exhibits larger variations than the QMI fit, and the two diverge in trend beyond~3.0\gev.

Possible additional contributions from the $D_{\squark J}(3040)^+$ state, with allowed $J^P$ values of $0^+$,~$1^-$, ~$2^+$ or $3^-$, are tested but found to be statistically marginal, yielding significance less than~$2\sigma$.
In the QMI method, the $J^P=1^-$ hypothesis is not tested as it is already implicitly included in the QMI component.
The statistical significances of contributions to the amplitude model are evaluated using Wilks' theorem~\cite{Wilks:1938dza}.
Although the $J^P$ of $D_{\squark J}(3040)^+$, which was observed in the \Dstar \kaon decays, remains unmeasured, prior angular analyses favor an unnatural spin-parity assignment~\cite{BaBar:2009rro}. Consequently, the $D_{\squark J}(3040)^+$ contribution is not included in the fit.
In addition, two nonresonant contributions are incorporated to describe the $\Dz\Kp$ $J^P = 0^+$ component: an exponentially decaying lineshape and a uniform lineshape. The exponentially decaying contribution describes a near-threshold enhancement that falls off slowly, while the uniform contribution accounts for a long tail extending above \mbox{3.0\gev}.
The statistical significance of the uniform contribution with respect to the model containing only the exponentially decaying contribution is approximately $3.9\sigma$.


A fit including only the known \Dz\Kp resonant contributions, hereafter referred to as the reference fit, describes the \Dz\Kp and \Dstarm\Kp spectra well, but fails to reproduce the \Dstarm\Dz distribution. 
A \chisq test over $64$ Dalitz‑plot bins yields $\chisq=82.1$, with the largest contributions concentrated in the \Dstarm\Dz region near 4.4\gev.
In that projection, a pronounced dip is observed in the data that cannot be explained by the \Dz\Kp amplitudes alone.

\begin{figure}
    \centering
    \includegraphics[width=\linewidth]{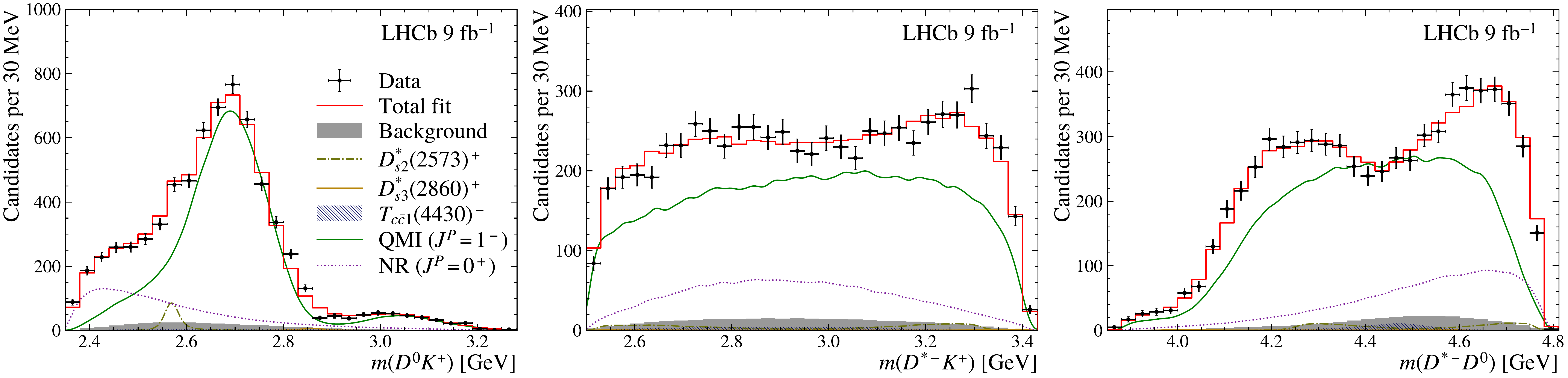}
    \caption{Distributions of the (left) $\Dz\Kp$, (middle) $\Dstarm\Kp$ and (right) $\Dstarm\Dz$ invariant masses from the baseline model including the \Zcm state.}
    \label{fig:baseline_fit}
\end{figure}

\begin{figure}[b!]
    \centering
    \includegraphics[width=0.6\linewidth]{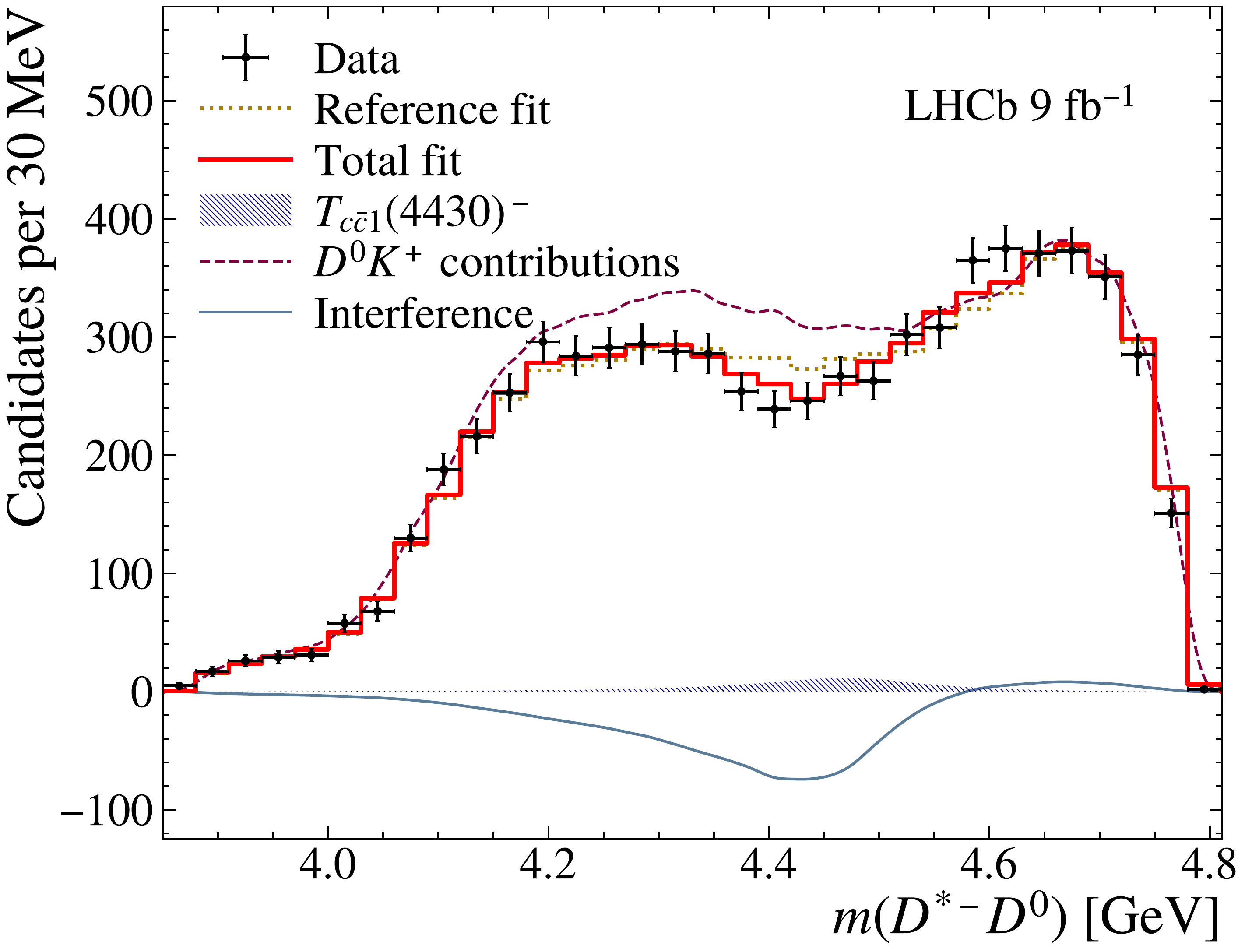}
    \caption{Distributions of the two-body invariant mass $m(\Dstarm\Dz)$. The total fit, \Dz\Kp contributions, and interference curves correspond to the baseline model including the \Zcm state, while the reference fit denotes the fit without the \Zcm. }
    \label{fig:4430fit}
\end{figure}

Given the proximity of the observed dip in the \Dstarm\Dz spectrum to the \Zcm mass, a relativistic Breit--Wigner function with mass and width fixed to its known values of $4478^{\,+\,15}_{\,-\,18}\mev$ and $181\pm31\mev$~\cite{PDG2024} is introduced to investigate the potential \Zcm contribution.
As the contribution of \Zcm is smaller than~2\%, only the lowest-energy $S$-wave component of the \decay{\Zcm}{\Dstarm\Dz} decay is considered. When including the $P$-wave decay of the \Zcm state, $2\Delta\text{NLL} = -5.1$ for $\Delta N_{\text{df}} = 2$, where $\Delta\text{NLL}$ denotes the change in the negative log-likelihood between two fits, and $\Delta N_{\rm df}$ the corresponding change in the number of degrees of freedom.
The fit projections of the baseline model including the \Zcm contributions are shown in Fig.~\ref{fig:baseline_fit}. A binned \chisq test over the same~64 Dalitz‑plot bins yields $\chisq = 72.8$, demonstrating an improvement over the fit without the \Zcm state. The $2\Delta\text{NLL}$ is $-43.0$ with $\Delta N_{\rm df}=2$, corresponding to a statistical significance of $6.2\sigma$. The comparison of the \Dstarm\Dz projection between fits with and without the \Zcm is shown in Fig.~\ref{fig:4430fit}.
The \Zcm fit fraction is~$(1.3\pm0.4)\%$, while its interference fit fractions with the QMI $S$-wave and the exponentially decaying nonresonant components are~$(-6.6 \pm 1.3)\%$ and~$(-2.0 \pm 0.5)\%$, respectively. This behavior is expected since large negative interference effects are needed to generate the observed dip structure.
Furthermore, the inclusion of a potential $\Dz\Kp$ state with $J^P=0^+$ does not alter the results for the \Zcm component.

To test consistency of the \Zcm mass and width with the known values~\cite{PDG2024}, a fit in which both parameters are allowed to vary freely is performed.
This yields an improvement of $2\Delta\text{NLL} = - 3.0$, with the fitted \Zcm mass and width being $4424 \pm 21 ^{\,+\,12}_{\,-\,6} \mev$ and $145 \pm 39 \pm 17\mev$, respectively, where the first uncertainty is statistical and the second is systematic.
The procedure used to evaluate the systematic uncertainties is described below.
The fitted mass and width differ from the known values in the $\psitwos\pi$ channel~\cite{PDG2024} by $1.8\sigma$ and~$0.7\sigma$, respectively, where the total uncertainties from both measurements are combined.
The quantum numbers $J^P$ of the \Zcm contribution are verified and $J^P = 1^+$ is preferred over the $0^-$, $1^-$, $2^\pm$, and $3^+$ assignments by more than $5\sigma$.


The $\Tccbar(4200)^-$ state shares the same $J^{P}$ quantum numbers as \Zcm and lies in a similar mass region of the \Dstarm\Dz spectrum. 
When incorporated into the fit model instead of the \Zcm, with its mass and width fixed to the known values~\cite{PDG2024}, an improvement of $2\Delta\text{NLL}=-16.0$ is found, corresponding to a significance of $3.6\sigma$, which is less than that obtained with the \Zcm.
When both resonances are included simultaneously, the $\Tccbar(4200)^-$ significance diminishes to $1.6\sigma$. The upper limit on its fit fraction, including both statistical and systematic uncertainties, is 1.9\% at 95\% confidence level.

Contributions from the $\Tccbar(3900)^-$ and $T^*_{\cquarkbar\squarkbar1}(2900)^0$ states are also tested, and 
the improvements in fit quality are marginal, with statistical significances of $1.3\sigma$ and $2.6\sigma$, respectively. The corresponding upper limits on their fit fractions, with both statistical and systematic uncertainties included, are $0.24\%$ and $0.32\%$ at $95\%$ confidence level.

Systematic uncertainties are assessed in four categories: signal and background yield determination, background and efficiency modeling, decay amplitude descriptions, and fit model.
Uncertainties in the yield extraction are evaluated by replacing the \Bz signal lineshape with a Gaussian function, the background exponential with a first-order Chebyshev polynomial, and assigning the resulting variations in the amplitude fit  as the systematic uncertainty. 
Efficiency uncertainties arising from particle identification, trigger calibration, and  simulation sample size are estimated by generating alternative efficiency maps through variations of the calibration inputs and propagation of their associated uncertainties.
The dominant systematic uncertainties originate from amplitude modeling, which is probed by varying the Blatt--Weisskopf radius from its nominal value of $3.0\gev^{-1}$ to $1.5\gev^{-1}$ and $4.5\gev^{-1}$, as well as by adjusting fixed parameters, such as the masses and widths of $D_{\squark2}^*(2573)^+$, $D_{\squark3}^*(2860)^+$, and \Zcm states, by $\pm 1 \sigma$ of  their known uncertainties~\cite{PDG2024}.
The spline points in the QMI model are shifted by $\pm10\mev$ to explore their impact. For the fit model, systematic uncertainties are evaluated by comparing the baseline results with alternative scenarios, including the use of a $K$-matrix model to describe the \Dz\Kp $J^P=1^-$ component, and the inclusion of a potential additional \Dz\Kp resonant state identified through a comprehensive scan over all physically allowed $J^P$ hypotheses.
The largest deviation from the baseline fit across these variations is assigned as the systematic uncertainty.
The total systematic uncertainty is obtained by adding all contributions in quadrature.
After fully accounting for all systematic uncertainties, the presence of the \Zcm state is confirmed at the $5.3\sigma$ level.

The product of $\decay{\Bz}{\Tccbar\Kp}$ and $\decay{\Tccbar}{\Dstarm\Dz}$ branching fractions is determined by combining the fit fraction measured in this analysis with $\mathcal{B}(\decay{\Bz}{\Dstarm \Dz \Kp})= (2.47\pm0.10\pm0.18)\times10^{-3}$~\cite{BaBar:2010tqo}, yielding ${(3.28\pm0.99^{\,+\,0.71}_{\,-\,1.58}\pm0.28)\times10^{-5}}$, where the uncertainties are statistical, systematic, and due to the uncertainty on $\mathcal{B}(\decay{\Bz}{\Dstarm\Dz\Kp})$.
Furthermore, using results from the \lhcb and \belle experiments~\cite{LHCb-PAPER-2014-014, Belle:2013shl}, the ratio of partial decay widths for the \Zcm is determined as 
\begin{equation}
    \begin{split}
    \Gamma(\decay{\Tccbar}{\Dstarm \Dz}) /\Gamma(\decay{\Tccbar}{\psitwos \pim}) &= 0.96\pm0.29
    ^{\,+\,0.22}_{\,-\,0.48}{}^{\,+\,0.57}_{\,-\,0.30}.
    \nonumber
    \end{split}
\end{equation}
The uncertainties are given in the same order as for the branching-fraction result, with the third uncertainty accounting for the total uncertainty from external measurements.


In conclusion, an amplitude analysis of the \decay{\Bz}{\Dstarm\Dz\Kp} decay reveals that contributions from the \Dz\Kp spectrum alone cannot adequately explain the observed dip near~4.4\gev in the \Dstarm\Dz spectrum.
Incorporating the exotic state \Zcm as a contributing component significantly improves the fit, with a statistical significance exceeding~$5\sigma$.
The spin-parity is firmly established as $J^P = 1^+$, and the measured mass and width are consistent with previous observations in the $\psitwos\pim$ channel~\cite{PDG2024}.
This analysis presents the first observation of the \Zcm in an open-charm decay channel.
The contributions from $\Tccbar(3900)^-$, $\Tccbar(4200)^-$, and $T^*_{\cquarkbar\squarkbar1}(2900)^0$ are also tested but found to be insignificant.
The measured ratio of the partial decay widths of the \Zcm to open- and hidden-charm final states is near unity, which disfavors the dominance of either. This provides important input for understanding its internal structure. Dedicated theoretical calculations are still lacking and are needed to interpret this ratio.

\section*{Acknowledgements}
%
%
\noindent We express our gratitude to our colleagues in the CERN
accelerator departments for the excellent performance of the LHC. We
thank the technical and administrative staff at the LHCb
institutes.
We acknowledge support from CERN and from the national agencies:
ARC (Australia);
CAPES, CNPq, FAPERJ and FINEP (Brazil); 
MOST and NSFC (China); 
CNRS/IN2P3 and CEA (France);  
BMFTR, DFG and MPG (Germany);
NKFIH (Hungary);              
INFN (Italy); 
NWO (Netherlands); 
MNiSW and NCN (Poland); 
MEC/IFA (Romania); 
MICIU and AEI (Spain);
SNSF and SER (Switzerland); 
NASU (Ukraine); 
STFC (United Kingdom); 
DOE NP and NSF (USA).
We acknowledge the computing resources that are provided by ARDC (Australia), 
CBPF (Brazil),
CERN, 
IHEP and LZU (China),
IN2P3 (France), 
KIT and DESY (Germany), 
INFN (Italy), 
SURF (Netherlands),
Polish WLCG (Poland),
IFIN-HH (Romania), 
PIC (Spain), CSCS (Switzerland), 
GridPP (United Kingdom),
and NSF (USA).  
We are indebted to the communities behind the multiple open-source
software packages on which we depend.
Individual groups or members have received support from
RTP (Australia), 
FWO Odysseus grant G0ASD25N (Belgium), 
Key Research Program of Frontier Sciences of CAS, CAS PIFI, CAS CCEPP (China); 
Minciencias (Colombia);
EPLANET, Marie Sk\l{}odowska-Curie Actions, ERC and NextGenerationEU (European Union);
A*MIDEX, ANR, IPhU and Labex P2IO, and R\'{e}gion Auvergne-Rh\^{o}ne-Alpes (France);
Alexander-von-Humboldt Foundation (Germany);
ICSC (Italy); 
Severo Ochoa and Mar\'ia de Maeztu Units of Excellence, GVA, XuntaGal, GENCAT, InTalent-Inditex and Prog.~Atracci\'on Talento CM (Spain);
the Leverhulme Trust, the Royal Society and UKRI (United Kingdom).

\clearpage

\section{End Matter}
\label{sec:Supplementary-App}

The knot positions for each wave, which are the QMI description of the $D^0K^+$ $J^P=1^-$ amplitude, are listed in Table~\ref{tab:knots}.
In each wave, the complex amplitudes of the first and last knots are fixed to $0$, and that of the knot at $m=2.7\gev$ is fixed to $1$. This fixes the overall phase and scale convention. All other knots are free complex parameters, giving 18 complex free nodes, equivalent to 36 real parameters in the QMI description.
A reduced number of nodes is chosen for higher partial waves, reflecting the expectation that their contributions diminish with increasing angular momentum.

\begin{figure}[!b]
    \centering
    \includegraphics[width=\linewidth]{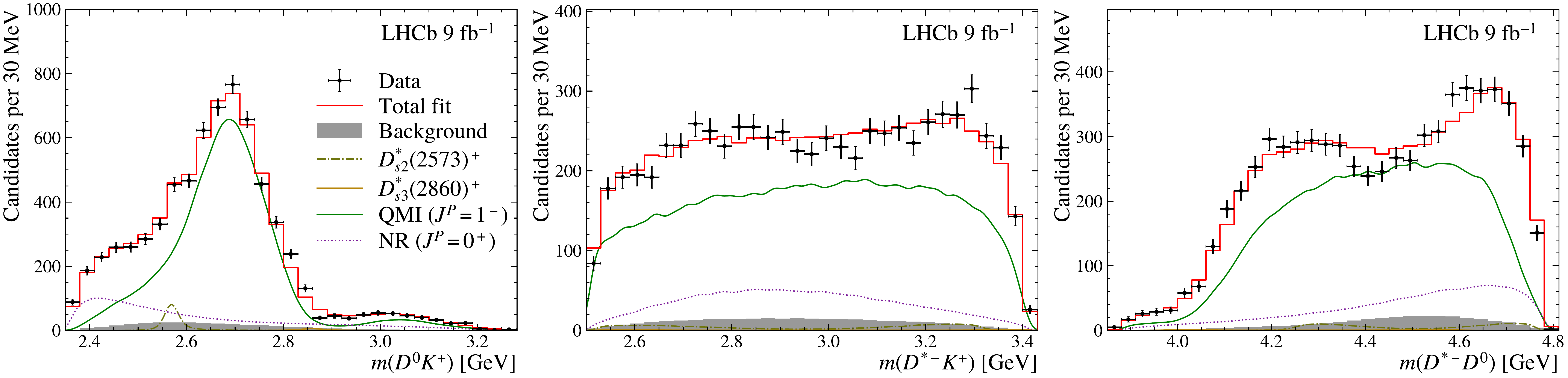}
    \caption{Distributions of (left) $\Dz\Kp$, (middle) $\Dstarm\Kp$, and (right) $\Dstarm\Dz$ invariant masses from the reference fit without the \Zcm state.}
    \label{fig:no4430}
\end{figure}

\begin{figure}
    \centering
    \includegraphics[width=0.45\linewidth]{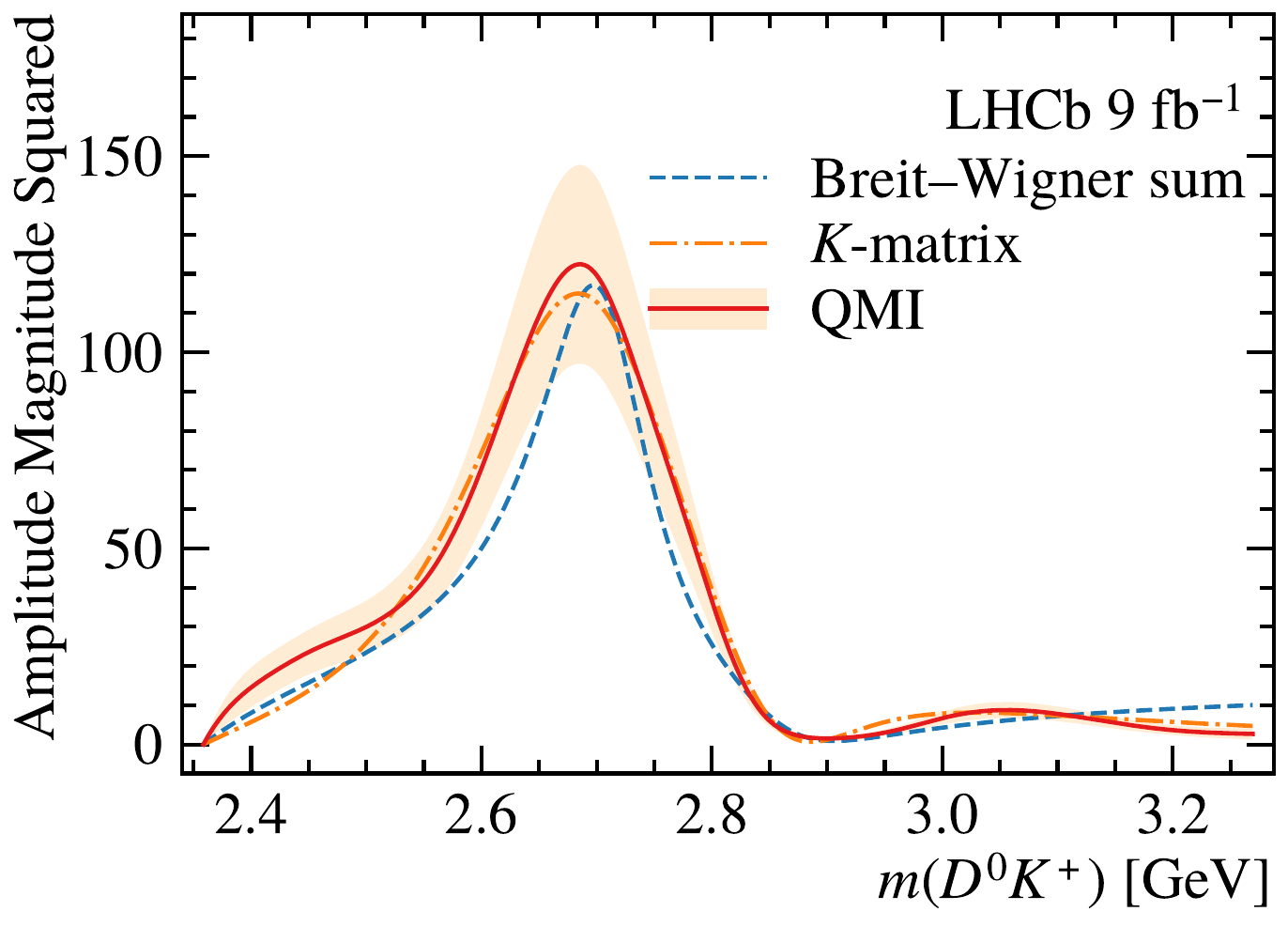}
    \includegraphics[width=0.47\linewidth]{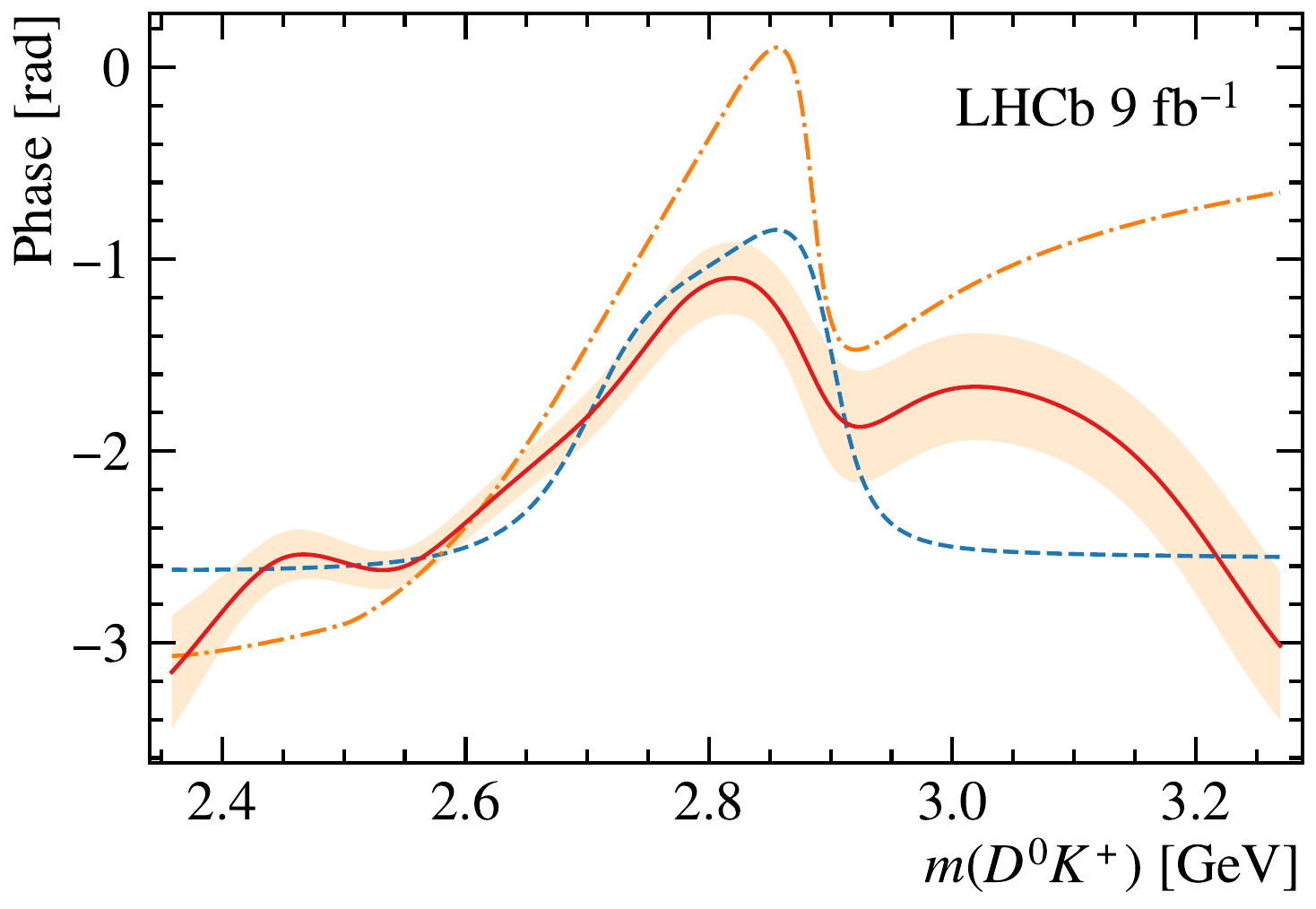}
    \caption{Squared magnitude of the amplitude and phase of the $L=0$ component of the $D^0K^+$ $J^P=1^-$ amplitude. The red line (with uncertainty band) corresponds to the QMI method, while the orange and blue dashed lines represent the $K$-matrix and Breit–Wigner sum models, respectively.}
    \label{fig:amp and phase}
\end{figure}

Figure~\ref{fig:no4430} shows the fit result of the reference model, which includes only the $D^0K$ spectrum.
Table~\ref{tab:fitfrac} shows the fit-fraction of the baseline model.
Figure~\ref{fig:amp and phase} shows the amplitude magnitude squared and phase of the $L=0$ component of the $D^0K^+$ $J^P=1^-$ amplitude, compared among the QMI, $K$-matrix and Breit--Wigner models. 

\begin{table}[!b]
\caption{Knot positions of the cubic-spline interpolation for the QMI description of the $D^0K^+$ $J^P=1^-$ amplitude.}
\begin{center}
\begin{tabular}{c|c}
\toprule
Partial wave & Knot positions [GeV] \\
\midrule
$S$-wave & 2.2, 2.35, 2.45, 2.55, 2.62, 2.7, 2.78, 2.85, 3.0, 3.2, 3.4 \\
$P$-wave & 2.2, 2.4, 2.5, 2.6, 2.7, 2.8, 3.0, 3.2, 3.4 \\
$D$-wave & 2.2, 2.4, 2.55, 2.7, 2.85, 3.05, 3.4 \\
\bottomrule
\end{tabular}
\end{center}
\label{tab:knots}
\end{table}

\begin{table}[!b]
\centering
\caption{
Fit fraction of each component in the QMI model with \Zcm, expressed in percent.
Here, $\mathrm{NR}_{e}$ and $\mathrm{NR}_{u}$ denote the exponentially
decaying and uniform nonresonant $D^{0}K^{+}$ components with
$J^{P}=0^{+}$, respectively.
The first and second uncertainties are statistical and systematic, with the latter excluding the fit-model uncertainty.
}
\label{tab:fitfrac}
\footnotesize
\setlength{\tabcolsep}{6pt}
\renewcommand{\arraystretch}{1.20}

\begin{tabular}{lc}
\toprule
Component & Fit fraction (\%) \\
\midrule
$D_{s2}^{*}(2573)^{+}$
    & $1.81 \,\pm\, 0.24 \,\pm\, 0.09$ \\
$D_{s3}^{*}(2860)^{+}$
    & $0.28 \,\pm\, 0.12 \,\pm\, 0.04$ \\
QMI $D$-wave
    & $1.35 \,\pm\, 0.23 \,\pm\, 0.10$ \\
QMI $P$-wave
    & $8.3 \,\pm\, 0.6 \,\pm\, 0.1$ \\
QMI $S$-wave
    & $66.6 \,\pm\, 1.4 \,\pm\, 1.9$ \\
$\mathrm{NR}_{e}$
    & $16.0 \,\pm\, 3.8 \,\pm\, 4.9$ \\
$\mathrm{NR}_{u}$
    & $3.0 \,\pm\, 1.8 \,\pm\, 2.5$ \\
\Zcm
    & $1.3 \,\pm\, 0.4 \,\pm\, 0.3$ \\
\bottomrule
\end{tabular}
\end{table}

\clearpage



\addcontentsline{toc}{section}{References}
\bibliographystyle{LHCb/LHCb}
\bibliography{main,LHCb/standard,LHCb/LHCb-PAPER,LHCb/LHCb-CONF,LHCb/LHCb-DP,LHCb/LHCb-TDR,LHCb/LHCb-PUB}

\newpage

\centerline
{\large\bf LHCb collaboration}
\begin
{flushleft}
\small
R.~Aaij$^{40}$\lhcborcid{0000-0003-0533-1952},
M.~Abdelfatah$^{72}$,
A.S.W.~Abdelmotteleb$^{60}$\lhcborcid{0000-0001-7905-0542},
C.~Abellan~Beteta$^{54}$\lhcborcid{0009-0009-0869-6798},
F.~Abudin\'en$^{62}$\lhcborcid{0000-0002-6737-3528},
T.~Ackernley$^{64}$\lhcborcid{0000-0002-5951-3498},
A.A.~Adefisoye$^{72}$\lhcborcid{0000-0003-2448-1550},
B.~Adeva$^{50}$\lhcborcid{0000-0001-9756-3712},
M.~Adinolfi$^{58}$\lhcborcid{0000-0002-1326-1264},
P.~Adlarson$^{88,45}$\lhcborcid{0000-0001-6280-3851}
% ...

{\footnotesize \it

% ...
$^{40}$Nikhef National Institute for Subatomic Physics, Amsterdam, Netherlands\\
% ...
$^{45}$National Center for Nuclear Research (NCBJ), Warsaw, Poland\\
% ...
$^{50}$Instituto Galego de F{\'\i}sica de Altas Enerx{\'\i}as (IGFAE), Universidade de Santiago de Compostela, Santiago de Compostela, Spain\\
% ...
$^{54}$Physik-Institut, Universit{\"a}t Z{\"u}rich, Z{\"u}rich, Switzerland\\
% ...
$^{58}$H.H. Wills Physics Laboratory, University of Bristol, Bristol, United Kingdom\\
% ...
$^{60}$Department of Physics, University of Warwick, Coventry, United Kingdom\\
% ...
$^{62}$School of Physics and Astronomy, University of Edinburgh, Edinburgh, United Kingdom\\
$^{63}$School of Physics and Astronomy, University of Glasgow, Glasgow, United Kingdom\\
$^{64}$Oliver Lodge Laboratory, University of Liverpool, Liverpool, United Kingdom\\
% ...
$^{72}$Syracuse University, Syracuse, NY, United States\\
% ...
$^{88}$Department of Physics and Astronomy, Uppsala University, Uppsala, Sweden, associated to $^{63}$\\
% ...
}
\end{flushleft}

\end{document}